\documentclass[journal=jctcce, manuscript=article]{achemso}
\usepackage{amsmath, amsfonts, amssymb, graphicx, subfigure, xcolor, soul, algorithm, algpseudocode}
\usepackage[labelfont=bf]{caption}

\author{Hao Tian}
\affiliation{Department of Chemistry, Center for Research Computing, Center for Drug Discovery, Design, and Delivery (CD4), Southern Methodist University, Dallas, Texas 75206, United States}

\author{Xi Jiang}
\affiliation{Department of Statistical Science, Southern Methodist University, Dallas, Texas 75206, United States}

\author{Sian Xiao}
\affiliation{Department of Chemistry, Center for Research Computing, Center for Drug Discovery, Design, and Delivery (CD4), Southern Methodist University, Dallas, Texas 75206, United States}

\author{Hunter La Force}
\affiliation{Department of Chemistry, Center for Research Computing, Center for Drug Discovery, Design, and Delivery (CD4), Southern Methodist University, Dallas, Texas 75206, United States}

\author{Eric C. Larson}
\affiliation{Department of Computer Science, Southern Methodist University, Dallas, Texas 75206, United States}

\author{Peng Tao}
\affiliation{Department of Chemistry, Center for Research Computing, Center for Drug Discovery, Design, and Delivery (CD4), Southern Methodist University, Dallas, Texas 75206, United States}
\email{ptao@smu.edu}

\title{LAST: Latent Space Assisted Adaptive Sampling for Protein Trajectories}

\begin{document}

\begin{tocentry}
\centering
\includegraphics[width=0.6\textwidth]{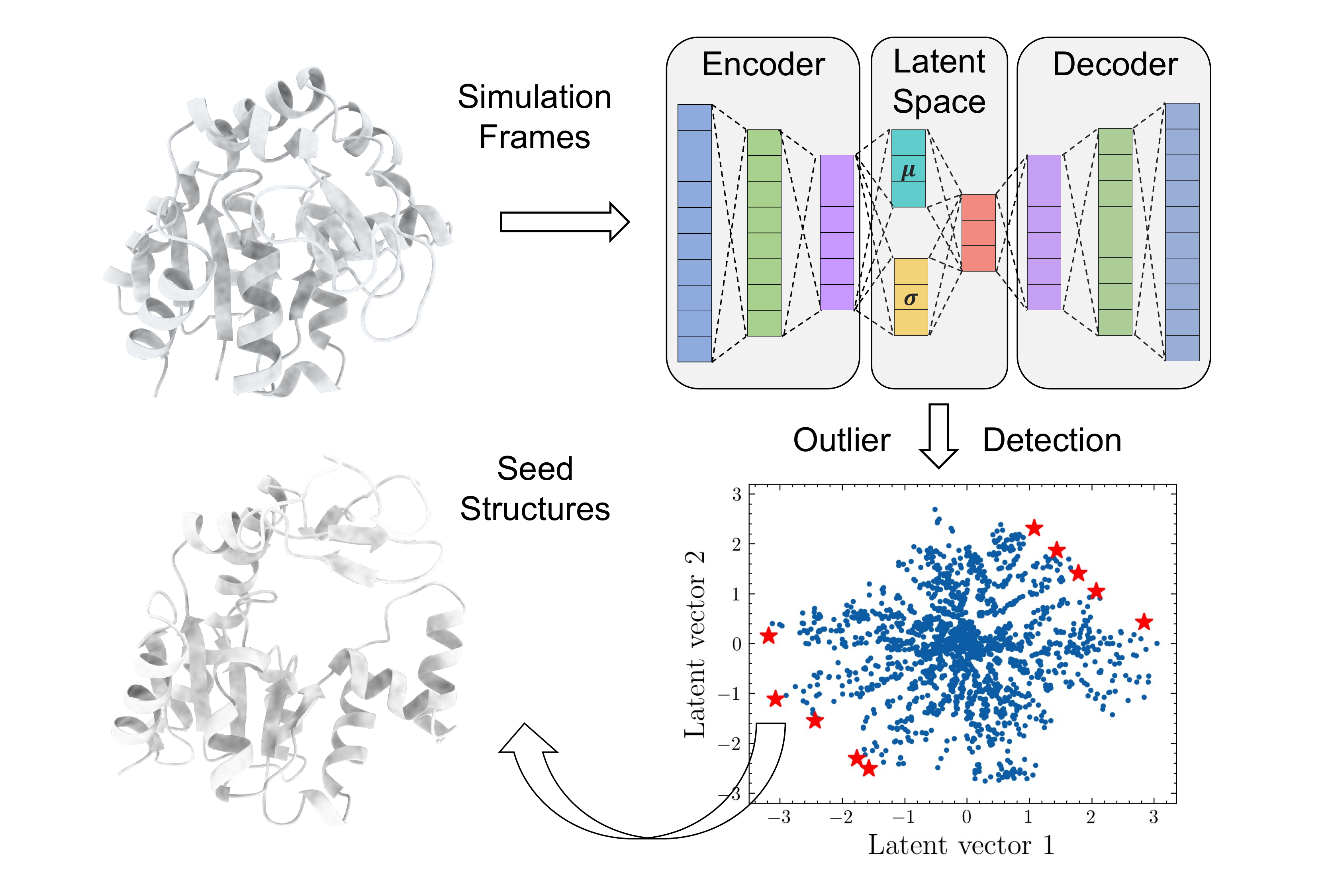}
\\

LAST: Latent Space Assisted Adaptive Sampling for Protein Trajectories

Authors: Hao Tian, Xi Jiang, Hunter La Force, Sian Xiao, Eric C. Larson and Peng Tao
\end{tocentry}

\clearpage

\begin{abstract}

Molecular dynamics (MD) simulation is widely used to study protein conformations and dynamics. However, conventional simulation suffers from being trapped in some local energy minima that are hard to escape. Thus, most computational time is spent sampling in the already visited regions. This leads to an inefficient sampling process and further hinders the exploration of protein movements in affordable simulation time. The advancement of deep learning provides new opportunities for protein sampling. Variational autoencoders are a class of deep learning models to learn a low-dimensional representation (referred to as the latent space) that can capture the key features of the input data. Based on this characteristic, we proposed a new adaptive sampling method, latent space assisted adaptive sampling for protein trajectories (LAST), to accelerate the exploration of protein conformational space. This method comprises cycles of (i) variational autoencoders training, (ii) seed structure selection on the latent space and (iii) conformational sampling through additional MD simulations. The proposed approach is validated through the sampling of four structures of two protein systems: two metastable states of \textit{E. Coli} adenosine kinase (ADK) and two native states of Vivid (VVD). In all four conformations, seed structures were shown to lie on the boundary of conformation distributions. Moreover, large conformational changes were observed in a shorter simulation time when compared with conventional MD (cMD) simulations in both systems. In metastable ADK simulations, LAST explored two transition paths toward two stable states while cMD became trapped in an energy basin. In VVD light state simulations, LAST was three times faster than cMD simulation with a similar conformational space.

\end{abstract}

\clearpage

\section{Introduction}

Molecular dynamics (MD) simulation has wide application on the study of protein conformations and dynamics. Recent developments in bio-computing, such as Anton \cite{shaw2009millisecond}, AMBER \cite{salomon2013routine} and OpenMM \cite{eastman2017openmm}, have enabled the simulation time scale to milliseconds, which promotes the research in sampling protein motions and structure landscapes \cite{prinz2011markov,lindert2013accelerated}. However, the time scales of many protein functions exceed the time scales achievable through traditional MD simulations. Moreover, protein sampling suffers from being trapped within local energy minima, proving difficult to escape. \cite{krivov2011free,brotzakis2018accelerating} As a result, most of the computational time is typically spent in sampling previously visited regions, which hinders efficient exploration of protein conformational space. 

Many enhanced sampling methods have been developed to address this issue. These methods can be classified into two types. In the first type, biasing potentials are introduced to expand the sampling space, such as metadynamics \cite{barducci2011metadynamics,raiteri2006efficient} and Gaussian-accelerated MD \cite{hamelberg2004accelerated}. In the second type, seed structures are selected as restarts for iterative MD simulations. This is referred to as adaptive sampling and numerous methods have been proposed that differ in seed selection methods. Markov state models have been applied to cluster conformations into microstates \cite{bowman2010enhanced}; parallel cascade selection MD (PaCS-MD) \cite{harada2013parallel} and nontargeted PaCS-MD \cite{harada2015nontargeted} calculates the root-mean-square deviation (RMSD) to select top snapshots; frontier expansion sampling \cite{zhang2020frontier} conducts dimensionality reduction with principal component analysis and Gaussian mixture models to select frontier structures.

Recent innovations in deep learning have provided new insights into sampling protein conformational space. \cite{chen2018molecular,hawkins2021generating} Autoencoders (AEs) and variational autoencoders (VAEs) are a class of deep learning models that learn a representation (encoding) which can capture the key features of input data. Several studies have demonstrated the success of AEs and VAEs in their applications to protein conformations and functions. \cite{ramaswamy2021deep,jin2021predicting,bandyopadhyay2021deep,guo2021generating} In our previous work \cite{tian2021explore}, we showed that VAEs are capable of learning a low-dimensional representation (referred to as the latent space) of protein systems. Through a quantitative study, the learned latent space is shown to be able to represent conformational characteristics. This property indicates that the larger differences two protein conformations have, the farther their corresponding latent points are from each other.

In this study, we proposed a new adaptive sampling method, latent space assisted adaptive sampling for protein trajectories (LAST), to accelerate the exploration of protein conformational space. This method iterates through three steps. First, a VAE is trained using previous MD simulations. Then, seed structures are selected on the learned latent space. Finally, additional simulations are conducted with these selected seeds. To quantify the performance, we applied LAST on four conformations in two protein systems: two metastable states of \textit{E. Coli} adenosine kinase (ADK) and two native states of Vivid (VVD). ADK conformations are projected onto two intrinsic angles while VVD conformations to RMSDs of two native structures. Our results showed that seed structures were consistently located on the boundary of sampled conformational distributions in all four conformations regardless of protein projection methods on reduced coordinates. We further compared the sampling efficiency between LAST and conventional MD (cMD). In both systems, large conformational changes were observed in a shorter time in LAST simulations. To be specific, LAST explored two transition paths toward two stable states while cMD being trapped in an energy basin in the metastable ADK simulations. In VVD simulations, LAST only took one third of cMD simulation time to discover a similar conformational space.

\section{Methods}

\subsection{Variational Autoencoder}

An autoencoder is a type of deep learning models that aims to encode a high-dimensional input to a low-dimensional latent space through an encoder module and decode it back to the original dimensions through a decoder module. By minimizing the differences between inputs and outputs, known as reconstruction loss, the latent space is expected to learn a low-dimensional representation of the input space. However, the latent space in an AE is not well-constrained and leads to unsatisfying results when sampling in the latent space. \cite{wetzel2017unsupervised} To overcome this issue, variational autoencoders add an optimization constraint on the latent space to follow a certain distribution. 

The encoder module $q_\phi(z|x)$ is an inference model that transforms data $x$ into output latent variable $z$, being parametrized with $\phi$. In reverse, the decoder module $p_\theta(x|z)$ is a generative model that transforms latent variable $z$ into output data $\hat{x}$, being parametrized with $\theta$. Both models are trained simultaneously with a joint distribution as $p(x, z) = p_\theta(x|z)p(z)$. $p(z)$ is the constraint distribution for latent space and typically is chosen as a normal distribution. \cite{doersch2016tutorial} The tractable variational Bayes approach is used to approximate the intractable posterior $p_\theta(z|x) = p_\theta(x|z)p(z)\mathbin{/}(\int{p_\theta(x|z)p(z)dz})$ by maximizing the Evidence Lower Bound (ELBO):

\begin{equation}
\mathcal{L}(\phi, \theta; x) = \mathbb{E}_{q_\phi(z|x)}[\log p_\theta(x|z)] - KL(q_\phi(z|x)||p(z)) \leq \log p_\theta(x)
\label{eq:elbo}
\end{equation}
where $KL$ is the Kullback-Leibler divergence. 

In our implementation, the VAE model is developed using Keras package \cite{chollet2015keras} with Tensorflow backend \cite{abadi2016tensorflow}. 

\subsection{Molecular Dynamics Simulations}
\label{section:md}

The initial structures of four conformations in two protein systems: two metastable states (PDB ID 1DVR and 2AK3) of \textit{E. Coli} adenosine kinase (ADK) and two native states (PDB ID 2PD7 and 3RH8) of Vivid (VVD) were taken from the Protein Data Bank (PDB) \cite{berman2000protein} . For each conformation, ligands and crystal waters were removed and chain A was extracted as the starting structure. The system was further solvated in a box of TIP3P water molecules and neutralized using Na$^+$ and Cl$^-$. Simulation files were generated using tleap \cite{case2005amber} with the AMBER ff14SB force filed \cite{maier2015ff14sb}. 100 ps NVT Langevin MD simulations were carried out, followed by 200 ps NPT simulations at 1 atm and 300 K. In each round of LAST method, one 100 ps MD simulation was conducted for each seed structure. Particle Mesh Ewald (PME) algorithm was used to calculate long-range electrostatic interactions. The simulation time step was set as 2 fs. All simulations were conducted with OpenMM 7 \cite{eastman2017openmm}.

\subsection{Latent Space Assisted Adaptive Sampling for Protein Trajectories}

\begin{figure*}[t]
  \includegraphics[width=0.9\textwidth]{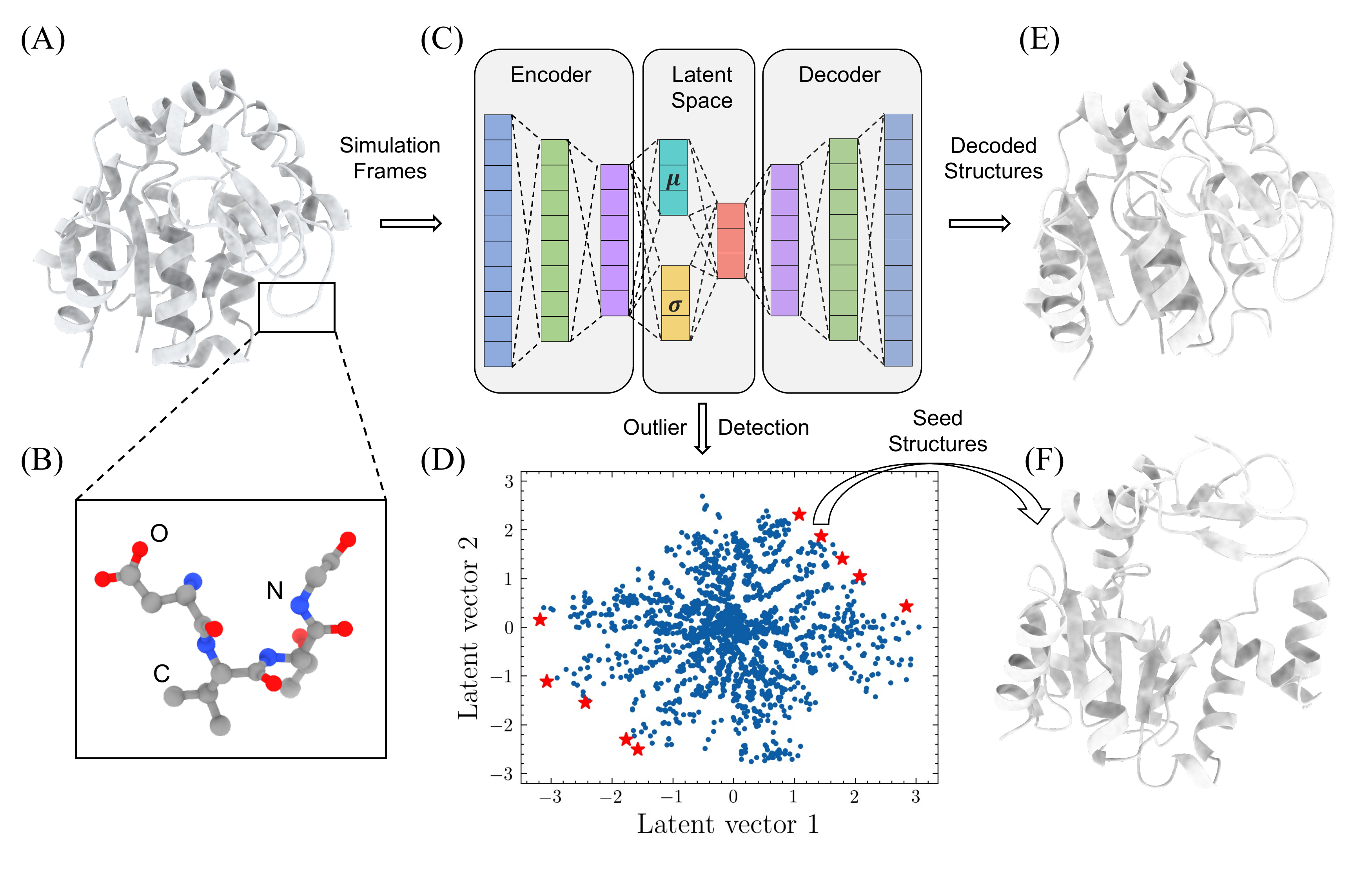}
  \caption{Workflow of LAST method. Heavy atoms in protein simulations are extracted as features to train a variational autoencoder model. In the latent space, outliers are selected based on the joint distribution of multivariate points in the latent space. Protein structures corresponding to these outliers are treated as seed structures to initiate additional simulations. LAST iterative conducts these steps until convergence. }
  \label{workflow}
\end{figure*}

LAST method includes three steps, and its workflow is shown in Figure \ref{workflow}. First, a variational autoencoder is trained using all previous simulations. Second, outliers are selected on the latent space and their corresponding protein structures are treated as seeds. Third, additional MD simulations are conducted from seed structures. 

\subsubsection*{VAE Training}

In each iteration, some preprocessing procedures are needed. The simulation trajectories are firstly aligned to the first frame and heavy atoms are extracted with Cartesian coordinates being expanded as a feature vector (Figure 1A-B). Then, each feature is transformed to a range of 0 to 1 using min-max linear scaling, which is used to construct a dataset for VAE training. 

The architecture of VAE model is shown in Figure \ref{workflow}C. In the current study, we design the encoder model being composed of three hidden layers with size of 512, 128 and 32 and decoder model with size of 32, 128 and 512. The number and size of hidden layers can be adjusted based on the size of proteins. The dimension of latent space is set as two for the simplicity and ease of visualization. 

\subsubsection*{Seeds Selection}

Appropriate seed selection method is needed to expedite the sampling of protein conformational space. In LAST, seeds are selected on the two-dimensional learned latent space of VAE, which has two important characteristics to enable an efficient seeds selection. First, as demonstrated in our previous work, the distance between two data points on the latent space is meaningful. Two structurally similar proteins have a shorter distance between their corresponding latent vectors. Second, the sampling distribution of latent space in the VAE is similar to a normal distribution due to the KL divergence term in the loss function. As for the distribution of the VAE latent space of protein conformations, the most common protein structures are encoded in the center of the latent space while structurally different proteins are encoded on the boundary. In a data distribution, outliers refer to those points that differ significantly from other data. Based on the above two points, it is reasonable to treat outliers on the latent space as seeds to accelerate conformational space exploration, as their conformations deviate the majority of the sampled ones. 

To implement the seed selection method, we propose three improvements to make LAST computationally efficient.

\begin{enumerate}
  \item Latent space of VAE is not strictly normal after optimization even though the normality is incentivized in the loss function. Therefore, a non-parametric multivariate kernel density estimator, instead of multivariate normal density function, is used to fit the latent space. The estimator is developed in Python statsmodels library \cite{seabold2010statsmodels}.
  \item Latent space distribution might be skewed so that the top N outliers with the smallest probability densities tend to gather on one side of distribution. To avoid the above issue, the cumulative distribution function (CDF) of combined data distribution on the latent space, instead of probability density, is applied to guarantee that samples from both sides of CDF (values close to 0 and 1) are equally selected. 
  \item Protein conformations corresponding to outliers can be located and selected based on data index. These protein conformations might be similar to each other, resulting in sampling repeated conformational space from MD simulations starting from these conformations. Thus, to further boost sampling efficiency, we require new seed structures to have at least 1$\mathring{A}$ RMSD with all previously selected seeds. 
\end{enumerate}

One example of seeds selection result is shown in Figure \ref{workflow}D, where seeds are highlighted in red stars in the latent space visualization. 

\subsubsection*{Additional MD Simulations}

Short MD simulations are conducted in each round. In the current study, 10 seeds are selected in one round and 100ps simulation is done starting from each seed. Thus, the total simulation time in each round is 1ns. The detail of these simulations is described in section \ref{section:md}.  

The above three steps are iteratively done until convergence. Here, we design the convergence criterion by calculating the mean RMSD of C$\alpha$ atoms with regard to the starting protein structure. The iterative sampling process is terminated once the mean RMSD stops to increase for successive five rounds or reaches the maximum round number. 

\begin{algorithm}
  \caption{Latent space assisted adaptive sampling for protein trajectories}
  \label{algo:last}
  \begin{algorithmic}
    \State Prepare simulation files.
    \State Conduct 100 ps NVT and 200 ps NPT simulations.
    \While{iteration is not reaching the maximum round}
      \State Align trajectories and extract Cartesian coordinates.
      \State Train a VAE model.
      \State Fit latent space with a non-parametric multivariate kernel density estimator.
      \State Select 10 outliers based on CDF and get seed structures.
      \State Conduct 100 ps simulation for each seed.
      \If{mean RMSD is converged}
        \State Stop iteration.
      \EndIf
    \EndWhile
  \end{algorithmic}
\end{algorithm}

The LAST algorithm is summarized in Algorithm \ref{algo:last} with codes are freely available at the GitHub site of https://github.com/smu-tao-group/LAST. 

\section{Results}

Four structures of two protein systems (ADK and VVD) were prepared for MD simulations as described in section \ref{section:md}. For each protein structure, 100 ps NVT and 200 ps NPT simulations were conducted. During the iterative process, all previous simulations were aligned to the first frame with Cartesian coordinates of heavy atoms being extracted as a feature vector to represent protein conformation. Afterwards, a variational autoencoder model was trained. 10 seed structures were selected with additional 100 ps simulation starting from each of them. Therefore, each iteration takes 1 ns simulation time. 

\begin{figure*}[t]
  \includegraphics[width=0.9\textwidth]{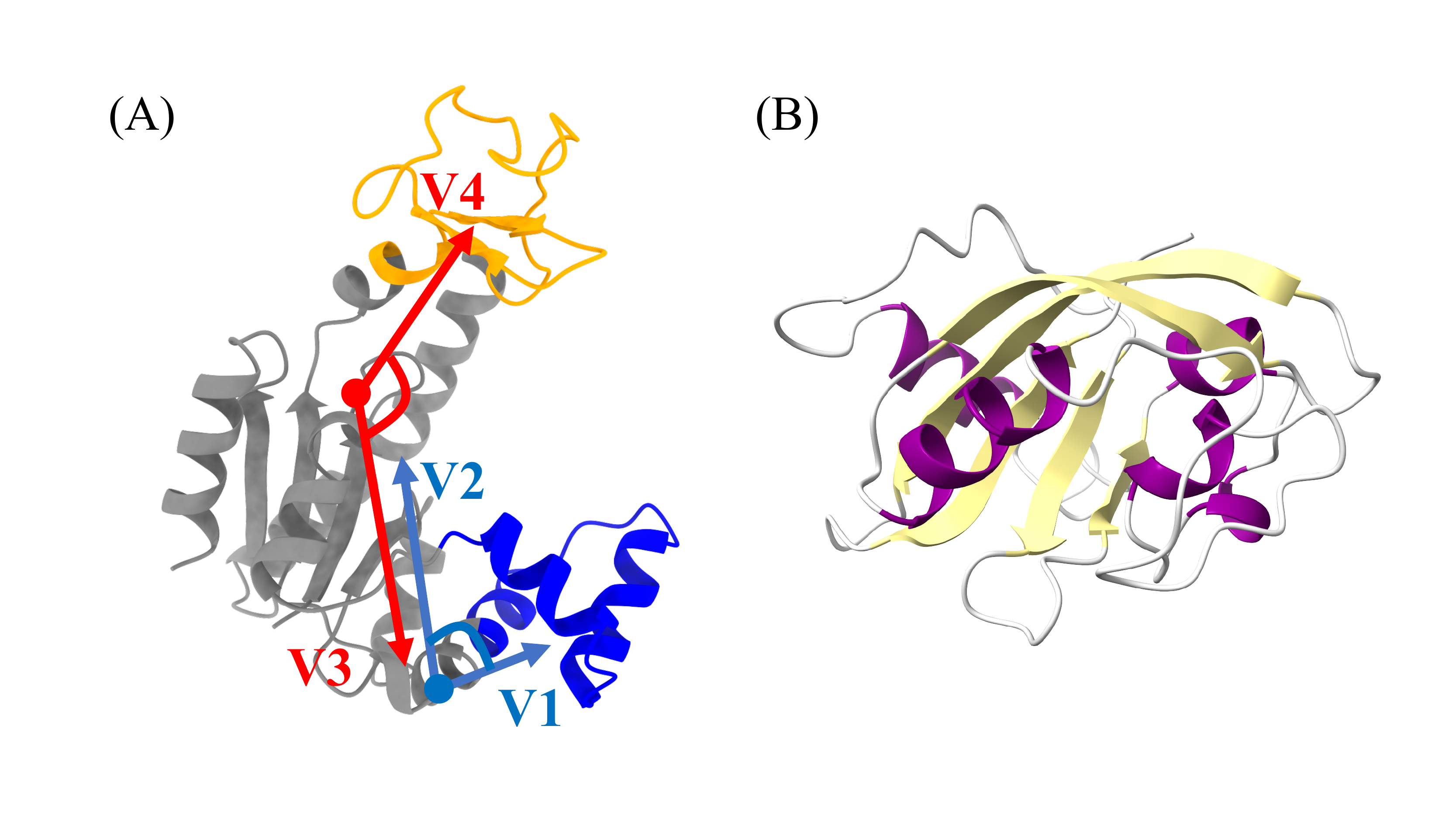}
  \caption{Structures of (A) ADK and (B) VVD. ADK is composed of a CORE domain (grey), a LID domain (orange) and a NMP domain (blue). LID-CORE and NMP-CORE angles are calculated by four vectors to represent protein conformations. VVD protein is colored at secondary structure level using VMD. }
  \label{vector}
\end{figure*}

ADK protein is composed of a rigid CORE domain, a lid-shaped ATP-binding domain (LID) and an AMP-binding domain (NMP). Many computational studies have shown ADK to carry out large conformational transitions between the closed state to the open state during the ATP-ADP catalyzation process. \cite{unan2015opening,matsunaga2012minimum} Four vectors that form NMP-CORE and LID-CORE angles have been widely used to characterize ADK protein conformation. VVD is a light-oxygen-voltage domain that undergoes global conformational changes upon perturbation. VVD is shown to be flexible in the native light state and relatively stable in the native dark state. \cite{matsunaga2012minimum} ADK and VVD structures are illustrated using ChimeraX \cite{goddard2018ucsf} (Figure \ref{vector}). 

Proper low-dimensional protein representations are needed to evaluate the quality of seed selection. In the current study, ADK protein structure is projected to LID-CORE and NMP-CORE 2D angle plot. We followed the same reside selection rule to calculate vectors and angles. \cite{tian2021explore} For VVD structure, 2D root-mean-square deviation (RMSD) with reference to the native dark and light structures was used to show the sampled protein conformational space. 

\begin{figure*}[t]
  \includegraphics[width=0.9\textwidth]{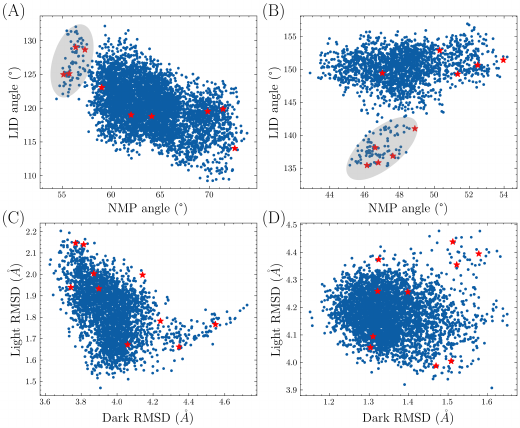}
  \caption{Seed structure distribution on low-dimensional protein representations. (A-B) ADK protein conformations are represented in LID-CORE and NMP-CORE angle vectors. (C-D) VVD protein conformation are represented in RMSDs with regard to the native dark and native light states. Seed structures are represented in red stars. Two less sampled regions are shown in grey circles.}
  \label{outlier}
\end{figure*}

Both the angle plot in ADK and RMSD plot in VVD were used to display the protein conformation of seed structures (Figure \ref{outlier}). In each subplot, seed structures are highlighted as red stars. In two metastable ADK conformations (Figure \ref{outlier}A-B), seed structures mainly locate in the less sampled regions with small or large LID/NMP angles. It should be noted that there are two significantly less sampled regions in both plots (grey regions in Figure \ref{outlier}A and B). 

This indicates that the variational autoencoder can capture the structural differences of protein conformations within the learned latent space. In the native dark and native light VVD conformations (Figure \ref{outlier}C-D), seed structures are also shown to be evenly distributed in the boundary of protein conformational space defined by RMSD to two native VVD structures. 

\begin{figure*}[t]
  \includegraphics[width=0.9\textwidth]{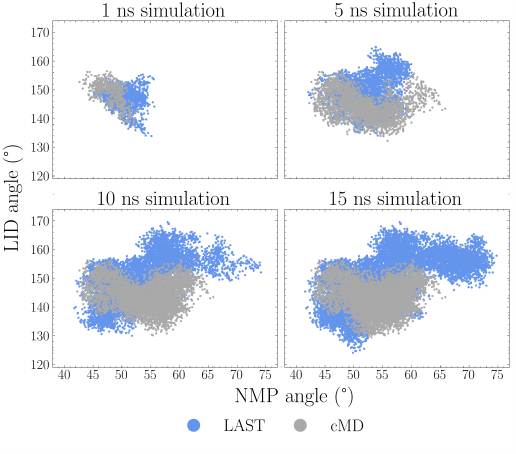}
  \caption{Comparison of ADK conformational spaces of LAST and cMD. Protein conformations are shown in blue at iteration 1, 5, 10 and 15 in LAST method. Protein conformations produced by cMD are shown in grey with the same simulation time. In each round LAST explored a larger conformational space compared with cMD. }
  \label{round}
\end{figure*}

To compare the effectiveness of LAST to conventional molecular dynamics simulations, the sampled protein conformational space in each round of LAST method is displayed together with cMD sampled conformations under the same simulation time. Figure \ref{round} shows the protein conformations in 1 ns, 5 ns, 10 ns and 15 ns for both LAST and cMD. Specifically, cMD was conducted twice independently. It is shown that under the same simulation time, LAST can explore more protein conformations than cMD. Moreover, the trained variational autoencoder can consistently learn a low-dimensional protein representation in the latent space, regardless of the growing number of simulations and changing shape of conformational space, and guide MD simulations to explore less sampled regions. In contrast, there are limited new conformations being explored in cMD simulations from 10 ns to 15 ns, indicating that it might be trapped in a local energy minimum. 

\begin{figure*}[t]
  \includegraphics[width=0.9\textwidth]{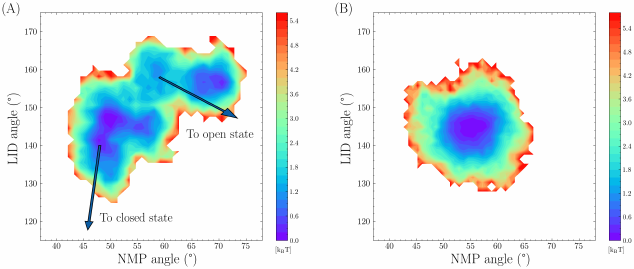}
  \caption{Free energy spaces of (A) LAST and (B) cMD produced protein conformations. LAST took 22 iterations to complete. Two independent 500 ns cMD were conducted for comparison. LAST explored two paths to the open and closed states while cMD being trapped in a local free energy minimum.}
  \label{energy}
\end{figure*}

We continued the LAST simulation of ADK until convergence. For comparison, two independent 500 ns cMD simulations were conducted. Both protein conformations with free energy being calculated are shown in Figure \ref{energy}. The LAST sampling method took 22 iterations (22 ns simulation time) and explored two paths from the metastable state to the two native states. This aligns with the computational finding that ADK protein undergoes conformational transitions between the open and the closed states. \cite{formoso2015energetics} Moreover, the ADK free energy landscape is similar to an Amber calculation of a previous study \cite{unan2015opening}, which the 2AK3 structure (NMP angle of 50$^{\circ}$ and LID angle of 155$^{\circ}$) is shown to lie in a less preferred state and quickly moves to other stable states. The two energy basins also agree with the free energy calculation results that the closed states are more stable than the open state. However, cMD simulation was being trapped in an energy basin, which is also indicated in the free energy plot of LAST. This demonstrates that LAST has a better chance to jump out of energy minimum without biasing potential energies. 

\begin{figure*}[t]
  \includegraphics[width=0.9\textwidth]{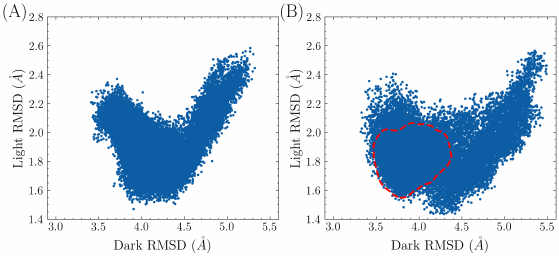}
  \caption{Comparison of VVD conformational spaces of (A) LAST and (B) cMD. LAST took 30 iterations to convergence. The conformational space in cMD with the same simulation time is circled on (B) with red dotted lines. cMD simulation was continued to 100 ns to have a similar space. }
  \label{3rh8}
\end{figure*}

For the VVD system, LAST simulation took 30 iterations (30 ns simulation time) to converge. The conformational space is illustrated in Figure \ref{3rh8}A. Two cMD simulations were independently conducted with the same simulation time. Their conformational space is highlighted as the red dotted circle in Figure \ref{3rh8}B. The region sampled by cMD overlaps greatly with the LAST sampled region but with small coverage of the overall conformational space. To compare the efficiency of two methods, these two cMD simulations were continued while this 2D RMSD map being monitored. It took 100 ns simulation time for cMD simulations to have a similar space shape with LAST. 

\begin{figure*}[t]
  \includegraphics[width=0.9\textwidth]{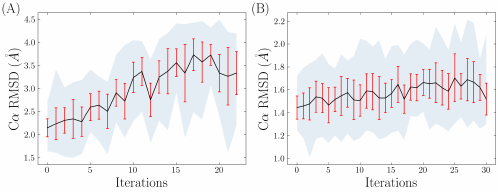}
  \caption{Mean RMSDs in (A) ADK and (B) VVD systems. Mean RMSD values are connected with black lines. Standard deviation in each iteration is plotted as vertical red lines. The gap between minimum and maximum RMSD values are colored as light grey background.}
  \label{rmsd}
\end{figure*}

The mean RMSDs with regard to the starting protein structure in each iteration were calculated for both systems and are shown in Figure \ref{rmsd}. Mean RMSDs are presented with black lines and standard deviation shown in red lines for each round. The maximum and minimum RMSD values are shown as the upper and lower bound in the colored regions. Currently, we set the patience as 5: the iteration loop stops if the maximum mean RMSD does not increase in 5 successive rounds. For the simulation in ADK system, RMSD starts with 2$\mathring{A}$, gradually increases to 3.5$\mathring{A}$, and stops at iteration 22. In contrast, the RMSDs in VVD system are smaller and the total simulation lasts longer with a total of 30 iterations. 

\section{Discussion}

In this study, we proposed a new adaptive sampling method to explore protein conformational space. LAST iteratively trains a VAE model using previous simulations, selects seeds that are structurally different to the sampled conformations, and uses them to initiate additional short MD simulations. LAST differs with previous methods in seed selection design, where outliers are selected and are treated as seeds on the latent space of VAE. VAE has been demonstrated to be effective in learning a low-dimensional protein representation in the latent space. \cite{hawkins2021generating,sultan2018transferable} The embeddings in the latent space are known to keep a distance similarity: if two protein structures are similar in structure, their embeddings in the latent space are close to each other. With this nature, the outliers on the latent space are worth further exploration through MD simulations, as their corresponding protein structures deviate from the most common structures. In LAST, these outliers are treated as seed structures to conduct additional MD simulations.  

The normality of latent space provides new opportunity for seed selection. However, the latent space does not strictly follow a normal distribution. This is mainly because of relatively strong emphasis on reconstruction loss and lesser emphasis on KL divergence during VAE training. The reconstruction loss term controls the quality of latent space data reconstruction (how well the VAE can reconstruct a protein structure) and KL divergence term constrains the distribution of the latent space (to what degree the latent space needs to follow a normal distribution). Therefore, in order to have a well-constructed and normal regularized latent space, appropriate weights are needed to be set for both terms. This is a challenging task with finetuning by hand, as the sample size keeps growing linearly with additional MD simulations in each round. Therefore, instead of trying to find weights to balance the reconstruction loss and KL divergence, we allow the latent space not strictly follow a normal distribution and use a non-parametric multivariate kernel density estimator to fit the latent space. 

One potential problem is that the distribution of the latent space might be skewed or kurtotic. In such cases, one side of probability density function will have a long tail with low values. This could lead to the situation that all selected seed structures lie in the long tail side, and the corresponding protein structures of these seeds might be similar to each other. Seeds gathering on one side of latent space distribution decreases the chance to explore more structurally different conformations and thus leads to a less efficient protein sampling process. To partially overcome this issue, we used the cumulative distribution function to select outliers: data points on the two sides of the CDF are evenly selected. This improvement prevents sampling similar seeds on the boundary of protein conformational spaces. 

Still, seed structures might be similar to each other. Nontargeted PaCS-MD proposed a nonredundant selection rule which calculates pairwise RMSDs between the current simulation cycle and seeds selected in all the past cycles. \cite{harada2019nontargeted} Protein configurations with large RMSD are then selected as new seeds in the current cycle. We took reference from this idea to select seeds. Outliers from two ends of the estimated CDF are picked sequentially while the pairwise RMSDs to previously selected seeds are calculated. We required the RMSD threshold should be greater than 1 $\mathring{A}$. If not, LAST discards this outlier and moves to the next. Moreover, LAST is a memory method: the selected seed structures are stored for RMSD calculation in future iterations, which avoids repeated sampling in the same conformational region and further improves the sampling efficiency. 

There are some tuning parameters in the LAST sampling scheme, including the number of seed structures, the RMSD threshold in seed selection, the architecture of VAE model and the number of rounds in convergence. In this work, 10 seeds are selected in each round. This could be changed under different protein systems and is subjected to the available computing resources. Also, the MD simulation time starting from seeds, currently set as 100 ps, can be adjusted accordingly. However, it should be noted that this simulation time should match the RMSD threshold: the simulation time should not be too short with large RMSD threshold. Given that the conformational space of selected seeds is not likely to be visited again, it is expected to have a reasonable simulation time to fully explore the conformations in each additional MD run. Besides, the number of hidden layers in VAE model is important to learn a useful latent space. Our previous finding suggests that a VAE model with three hidden layers is sufficient to learn the ADK protein conformations. Larger model architectures do not have a significant improvement but instead will lead to longer training time. The proper architecture of VAE, in terms of the number of hidden layers and the number of dimensions in the latent space, is worth studying to provide general guidelines when dealing with different protein families. Lastly, it is worth noting that the convergence criterion used in this study does not represent the "true" convergence of protein systems. The notion of "true" convergence, as discussed in previous studies, \cite{romo2011block,sawle2016convergence,knapp2011intuitive} is elusive in simulations. More appropriate criteria are needed for the convergence signal in adaptive sampling, through either numerical indicators or self-consistency checks. 

\section{Conclusion}

In this study, we present an adaptive sampling method, latent space assisted adaptive sampling for protein trajectories, to accelerate the exploration of protein conformational spaces. LAST iterates through variational autoencoder training, seed selection and additional short MD simulations. LAST differs with previous methods in seed selection where the outliers in the learned latent space are selected and treated as seed structures. We tested LAST method in ADK and VVD protein systems, each with different low-dimensional protein representations. In both systems, LAST can capture the key protein characteristics and select seeds that lie in the boundary of conformational space. For ADK simulations, LAST explored two transition paths that are in agreement with previous findings. For VVD simulations, LAST is three times faster than conventional MD for exploring the same conformational regions.

\begin{acknowledgement}

Research reported in this paper was supported by the National Institute of General Medical Sciences of the National Institutes of Health under Award No. R15GM122013. Computational time was generously provided by Southern Methodist University's Center for Research Computing.

\end{acknowledgement}

\bibliography{ref-abbr, ref}


\end{document}